\documentclass[reprint,aps,prl,twocolumn,
               notitlepage,mathrsfs,
               amssymb,amsmath,amsfonts,
               superscriptaddress,
               longbibliography,
               nofootinbib,floatfix]{revtex4-2}

\usepackage{graphicx}
\usepackage{dcolumn}
\usepackage{bm}
\usepackage{mathtools}

\usepackage{chemformula}

\usepackage[dvipsnames]{xcolor}

\definecolor{BerlinU1}{HTML}{62AD2D}
\definecolor{BerlinU2}{HTML}{E94D10}
\definecolor{BerlinU3}{HTML}{00A192}
\definecolor{BerlinU9}{HTML}{F18800}

\usepackage[colorlinks=true]{hyperref}
\hypersetup{
    colorlinks = true,
    linkcolor = BerlinU9,
    urlcolor = BerlinU9,
    linkbordercolor = {white},
    citecolor = BerlinU9,
    }

\usepackage{orcidlink}
\begin{document}

\title{Enhancing eccentricity through tidal resonance in binary neutron star systems}

\author{Ivan \surname{Markin}\,\orcidlink{0000-0001-5731-1633}}
\email{ivan.markin@uni-potsdam.de}
\affiliation{Institut f\"ur Physik und Astronomie, Universit\"at Potsdam, Haus 28, Karl-Liebknecht-Str. 24/25, 14476, Potsdam, Germany}

\author{Natalie \surname{Williams}\,\orcidlink{0000-0002-5656-8119}}
\affiliation{Institut f\"ur Physik und Astronomie, Universit\"at Potsdam, Haus 28, Karl-Liebknecht-Str. 24/25, 14476, Potsdam, Germany}

\author{Tim \surname{Dietrich}\,\orcidlink{0000-0003-2374-307X}}
\affiliation{Institut f\"ur Physik und Astronomie, Universit\"at Potsdam, Haus 28, Karl-Liebknecht-Str. 24/25, 14476, Potsdam, Germany}
\affiliation{Max Planck Institute for Gravitational Physics (Albert Einstein Institute), Am M\"uhlenberg 1, Potsdam 14476, Germany}

\date{\today}

\begin{abstract}
Orbital eccentricity of binary neutron star systems is a key parameter to reveal their formation history. Since binaries formed in isolation are expected to circularize long before merger, measurable eccentricity suggests dynamical formation in dense stellar environments. However, using numerical-relativity simulations, we show that eccentricity can be introduced in the late inspiral of an initially quasicircular binary, due to the $f$-mode resonance of one of the stars. We simulate the merger of two highly spinning neutron stars, one star spinning prograde and the other retrograde. Persistent tidal resonance with the $f$ mode of the retrograde-spinning star is observed, accompanied by a gradual change of its spin due to angular momentum transfer from the orbit. This spin change can be well explained by a simple resonance locking model that accounts for the spin-dependent shift of the $f$-mode frequency. These findings reveal an origin of eccentricity in binary neutron stars that is independent of their formation history and provide important information for modeling binary neutron star dynamics with rapidly spinning stars.
\end{abstract}

\maketitle

\paragraph{\textbf{Introduction---}}
The gravitational-wave (GW) signals from compact binary coalescences, such as binary neutron stars (BNS) formed via isolated binary evolution~\cite{Paczynski1976,Ostriker1976,Bethe:1998bn,Belczynski:2001uc}, are expected to enter the sensitivity band of ground-based GW detectors with vanishing orbital eccentricity~\cite{Peters:1964zz,1977A&A....57..383Z,Belczynski:2001uc,Kowalska:2010qg,Huerta:2013qb}. Therefore, a confident detection of nonzero eccentricity in the GW signals generally suggests their dynamical formation history in dense stellar environments, such as globular and nuclear clusters~\cite{Samsing:2017xmd,Rodriguez:2017pec,Fragione:2018yrb,Zevin:2018kzq} or young stellar clusters~\cite{Banerjee:2016ths,Banerjee:2017mgr,DiCarlo:2019pmf}.
Notably, due to the rapid progress in the development of eccentric GW models, reanalyses of a black hole-neutron star merger signal GW200105\_162426~\cite{LIGOScientific:2021qlt} showed support for nonzero eccentricity~\cite{Morras:2025xfu,Planas:2025plq,Jan:2025fps,Kacanja:2025kpr}, although this might be prior-driven~\cite{Clarke:2026cuw}.

Eccentricity may also be acquired after formation. For example, in quasicircular systems, it can be introduced by a third body due to mechanisms such as the von~Zeipel-Lidov-Kozai effect~\cite{Zeipel:1909,Lidov:1962,Kozai:1962} (cf.~also~Ref.~\cite{Heinze:2025usf} for numerical-relativity simulations where eccentricity is introduced to a quasicircular black hole binary by a third black hole).

Alternatively, orbital eccentricity of a binary can also be introduced due to a sudden or continuous change of the binary component properties, particularly if the change is asymmetric.
Such changes can occur due to external forces, e.g., the drag force of the surrounding gas in which the binary is immersed can remove angular momentum from a binary and increase its eccentricity~\cite{Armitage:2005xq,Cuadra:2008xn,Ishibashi:2020zzy,Cardoso:2020iji}. Also, a drag force from dark matter can significantly boost eccentricity in intermediate-mass-ratio binary black hole (BBH) inspirals~\cite{Yue:2019ozq,Becker:2021ivq}.
Such changes can also occur in the intrinsic properties of the components. For example, if a star in a binary system experiences mass loss, the orbital parameters would change, increasing orbital eccentricity via the Boersma-Blaauw recoil mechanism~\cite{1961BAN....15..291B,1961BAN....15..265B}. 
A neutron star (NS) in a binary can undergo a phase transition, inducing orbital eccentricity~\cite{Jiang:2015gpa,Alvarez-Castillo:2019apz}.
For a BBH merger in scalar Gauss-Bonnet gravity with the black holes having opposite scalar charges, one of the black holes can spontaneously flip its scalar charge sign and introduce orbital eccentricity~\cite{Lara:2025kzj}.

Another example of altering the state of the NS in a binary system is tidal resonance, where angular momentum and energy are transferred from the orbit to the star. Among the possible oscillation modes capable of undergoing resonance, the fundamental ($f$) mode has the strongest coupling with the orbit~\cite{Reisenegger:1994}. 
Tidal resonance with an $f$ mode occurs when the frequency of the mode matches the GW frequency or, equivalently, the angular frequency of the mode pattern matches the orbital angular frequency~\cite{Lai:1993di,Kokkotas:1995xe,Lai:1997wh,Ho:1998hq}. For nonspinning NSs, the resonance is often unattainable during the inspiral as the $f$-mode frequency lies beyond merger, at  $\sim1.5-2.2$~kHz~\cite{1983ApJS...53...73L,Andersson:1997rn} for NSs with masses $\sim 1.4~\mathrm{M_\odot}$. 
In the case that the NS is spinning, the $f$ mode splits into co- and counterrotating modes~\cite{Ledoux:1951,Robe:1968,Gaertig:2008uz}, with the frequency of the corotating mode increasing, and the frequency of the counterrotating mode decreasing. This means that for an NS with its spin antialigned with the orbital angular momentum, tidal resonance of the orbit with its counterrotating mode may be reached during late inspiral.

Ref.~\cite{Kuan:2024jnw} presented the first numerical-relativity simulations of such a BNS system, with high NS spins antialigned with the orbital angular momentum. There, the tidal resonance occurs for both NSs at the same time and additionally changes the spins of both NSs. Importantly, the binary remains quasicircular even after undergoing resonance. This is likely a consequence of the complete symmetry of the binary (equal masses and spins), which would result in orbital readjustments occurring simultaneously and preserve circularity.
However, this does not need to hold once the binary has an asymmetry of some kind.

In this \textit{Letter}, we demonstrate, for the first time, that orbital eccentricity can be introduced during the late inspiral of a BNS system where only one NS undergoes tidal resonance. We show this by performing numerical-relativity simulations of a BNS system where the spins of its components are antialigned to each other, and parallel to the angular momentum of the binary.

We also propose a simple model for the evolution of the resonating NS spin under the assumption of resonance locking. The excellent agreement of the model with the simulation data supports the finding that once tidal resonance sets in, the resonance remains active until merger via a feedback loop due to the spindown of the resonating NS.
Accounting for this scenario in waveform models can enable a more accurate description of such resonating BNS systems and help constrain properties of NS matter.

Throughout the \textit{Letter}, unless stated otherwise, we use geometrized units $G=c=\mathrm{M_\odot}=1$.

\paragraph{\textbf{Physical setup---}}
We simulate an equal-mass binary with gravitational NS masses of $m_1=m_2=1.35~\mathrm{M_\odot}$ described with a piecewise-polytropic representation of the SLy equation of state~\cite{Douchin:2001sv,Read:2008iy}, and individual tidal deformabilities $\Lambda_1=\Lambda_2=414.4$.
Both NSs have an initial dimensionless spin magnitude $|\chi|=0.5$ (corresponding to a rotation frequency of $895.5$~Hz), with the first NS spin aligned with the orbital angular momentum, and the second antialigned. We simulate the binary starting from the orbital frequency $163.4$~Hz to allow the binary to evolve in the quasicircular, preresonance regime. We employ \textsc{FUKA}~\cite{Grandclement:2009ju,Papenfort:2021hod} for the construction of the initial data, and perform eccentricity reduction following Refs.~\cite{Pfeiffer:2007yz,Papenfort:2021hod}. We evolve the binary using \textsc{BAM}~\cite{Bruegmann:2006ulg,Thierfelder:2011yi,Dietrich:2015iva,Bernuzzi:2016pie,Dietrich:2018phi} at two resolutions: R192 ($145$m grid spacing) and R128 ($217$m grid spacing). Unless stated otherwise, we refer to the results of the higher resolution, R192. The numerical setup is inherited from Ref.~\cite{Kuan:2025bzu}, to which we refer the reader for further details.

A key quantity we monitor throughout the simulations is the local spin indicator $\hat{\chi}$ for each NS, which mimics the dimensionless spin parameter $\chi$ and follows the convention of Eq. (59) of Ref.~\cite{Lam:2022yeg}, namely,

\begin{equation}
\begin{split}
    \hat{\chi} = \frac{1}{m^2} & \int d^3x \sqrt{\gamma}  D~\times  \\ &h \left[ (u_y - \bar{u}_y)(x-x_c)  -  (u_x - \bar{u}_x)(y-y_c)\right],
\end{split}
\end{equation}

where $m$ is the gravitational mass of the NS in isolation at the start of the simulation, $\gamma$ is the determinant of the spatial metric, $D=\rho W$ is the conserved rest-mass density, $h$ is the specific enthalpy, $u_{x,y}$ are the components of the fluid four-velocity along $x$ and $y$ axes, respectively, and $\bar{u}_{x,y}$ are the averages across the control sphere, $x$ and $y$ are the coordinates of the fluid element, and $x_c,y_c$ are the coordinates of the NS center defined here as the point of the minimum lapse. The integral is taken over the control sphere, which is centered around the NS and has a coordinate radius of $12~\mathrm{M_\odot}$. The NS gravitational mass is assumed to be constant throughout the inspiral, which is motivated by the conservation of the baryonic mass. Hence, $\hat{\chi}$ only accounts for the changes in the angular momentum of the NS.

\begin{figure}[!tp]
    \centering
    \includegraphics[width=1\linewidth]{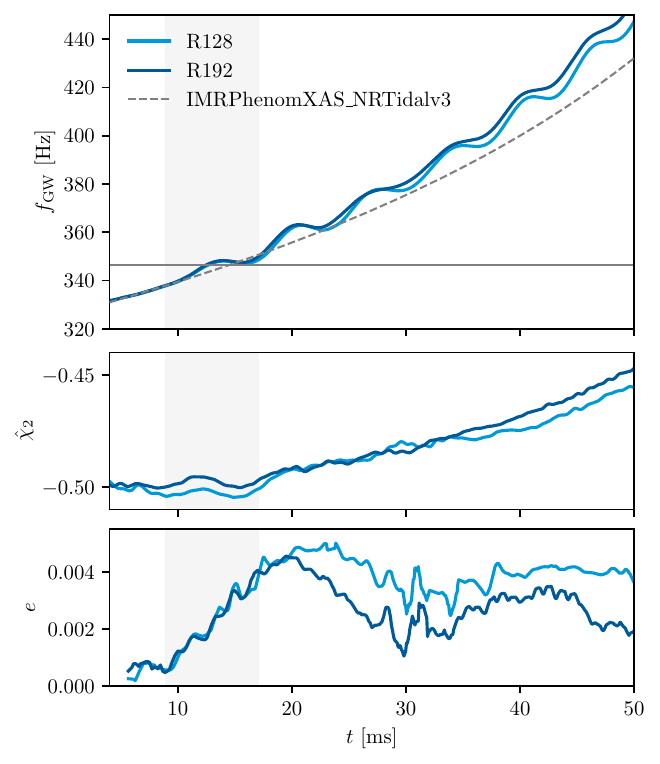}
    \caption{\textit{Top:} Evolution of the GW frequency over time for two resolutions, higher R192 and lower R128. The horizontal solid gray line indicates the frequency of the onset of the tidal resonance locking defined by the continuous spin change of the resonating NS. 
    The dashed gray line shows the evolution of the frequency predicted by the \mbox{IMRPhenomXAS\_NRTidalv3} model for our configuration.
    \textit{Middle:} Evolution of the dimensionless spin parameter for the resonating NS $\hat{\chi}_2$ over time.
     \textit{Bottom:} Evolution of the orbital eccentricity over time. The data for the first $\sim 5.5$~ms is not shown as the proper distance data is contaminated by gauge effects.
     \textit{All panels:} The light-gray vertical band represents the adjustment window during which most eccentricity is induced.
     } 
    \label{fig:decspin_main_plot_stacked}
\end{figure}

\paragraph{\textbf{Introduction of the eccentricity---}}
In Fig.~\ref{fig:decspin_main_plot_stacked}, we present the time evolution of the GW frequency $f_\mathrm{GW}$, spin of the resonating NS $\hat{\chi}_2$, and the eccentricity $e$, measured from the proper distance time series over a sliding window $7.5$~ms wide.
In the absence of resonance, the spins of both NSs would remain constant. However, due to the tidal resonance with the counterrotating $f$ mode of the NS with negative spin, orbital angular momentum is transferred to the NS, reducing the magnitude of its spin. Similar behavior was first reported in Ref.~\cite{Kuan:2024jnw}, where both NSs undergo resonance and spin down simultaneously.
Here, due to the asymmetric spin orientation, only one NS undergoes the resonance.
For the first $\sim9$~ms, the GW frequency follows the quasicircular evolution predicted by \mbox{IMRPhenomXAS\_NRTidalv3}~\cite{Abac:2023ujg}. Afterwards, it shows signs of eccentricity and evolves faster than predicted by the waveform model.

We refer to the interval $\sim9-17$~ms as the \textit{adjustment window}, during which the resonance causes the orbit to adjust, and orbital eccentricity is induced.
First, at $\sim9-12$~ms, the spin of the resonating NS $\hat{\chi_2}$ begins to increase, marking the onset of the angular momentum transfer to the NS. Then, from about $\sim 12-16$~ms, the spin of the resonating NS reduces back to approximately the initial value.
Throughout the adjustment window, the eccentricity grows roughly linearly from $e \sim 7 \times 10^{-4}$ to $e \sim 4 \times 10^{-3}$ for both resolutions.
Afterwards, the eccentricity temporarily drops to $e \sim 2 \times 10^{-3}$ ($e \sim 3 \times 10^{-3}$ for R128), and then grows again to $e \sim 3 \times 10^{-3}$ ($e \sim 4 \times 10^{-3}$ for R128). We suggest that the evolution of the eccentricity during that time is a result of the nonlinear interplay between the radiation of angular momentum via GWs, the transfer of angular momentum to the NS, and the readjustments of the orbital energy.
Using the Peters formula~\cite{Peters:1964zz}, we find that the eccentricity of a nonresonating binary would need to be $e_\mathrm{20Hz} \sim 0.1$ at 20~Hz to decay to the values of the induced eccentricity in our simulations. This is close to the eccentricity inferred for  GW200105\_162426~\cite{LIGOScientific:2021qlt,Morras:2025xfu,Planas:2025plq,Jan:2025fps,Kacanja:2025kpr}.

One could attempt to estimate the induced eccentricity, at least as a first-order approximation, by introducing a rapid change of the angular momentum in a circular Keplerian orbit due to its transfer to the NS. The eccentricity for a Keplerian orbit is
\begin{equation}
    e = \sqrt{1 + \frac{2}{\mathcal{M}_c^5} L^2 E},
\end{equation}
where $\mathcal{M}_c=(m_1 m_2)^{3/5}/(m_1+m_2)^{1/5}$ is the chirp mass of the binary, $L$ is the orbital angular momentum, $E$ is the orbital energy. Now, introducing a small change in the angular momentum $\Delta L$ and in energy $\Delta E$ to a quasicircular orbit, one arrives at the induced eccentricity
\begin{equation}
    \label{eq:induced_eccentricity}
    e \sim \sqrt{2\frac{\Delta L}{L}+\frac{\Delta E}{E}}.
\end{equation}

For our case within the adjustment window, however, the resonance-related change of the angular momentum is not sudden, but gradual, and not always monotonic. In addition, at $f_\mathrm{GW}\sim346$~Hz, GW radiation carries away in the order of ten times more orbital momentum than is removed due to the resonance, which can affect the angular momentum redistribution. It is also not clear how the orbital energy readjusts during this process. 
Crucially, the model of Eq.~\eqref{eq:induced_eccentricity} fails to explain the preservation of the quasicircularity for a binary when both NSs resonate~\cite{Kuan:2024jnw}, as it is insensitive to how the angular momentum is removed. A more reliable description would require modeling the orbital change consistently and from first principles~\cite{Yue:2019ozq,Becker:2021ivq,Yu:2025ptm,Ghosh:2025ban}.

We note that the continuous angular momentum removal is analogous to an external dissipative force acting on one NS, which is known to induce high eccentricities in intermediate-mass-ratio BBH systems~\cite{Yue:2019ozq,Becker:2021ivq}.

\paragraph{\textbf{Tidal resonance locking---}}
The tidal resonance with the counterrotating $f$ mode occurs when the GW frequency reaches the mode frequency, $\sigma_i^\mathrm{u}$. 
If the mode frequency were constant, it would be expected that the resonance and angular momentum transfer would occur only over a narrow GW frequency band, leaving the NS in a perturbed state and slightly reduced spin. However, instead, the spin of the resonating NS continues to increase until merger, cf.~Fig.~\ref{fig:decspin_main_plot_stacked}. This behavior was first shown in Ref.~\cite{Kuan:2024jnw}.

\begin{figure}[tp]
    \centering
    \includegraphics[width=1\linewidth]{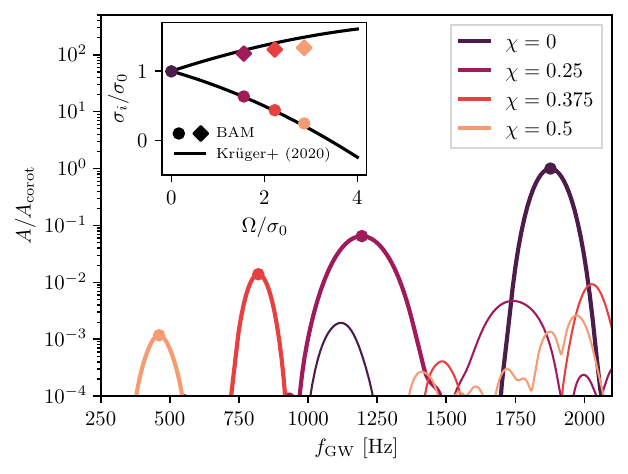}
    \caption{GW spectra of isolated spinning NSs with different spins, perturbed to excite the $f$ modes. The inset plot shows the quasiuniversal relation between the spin angular frequency and frequency of the co- and counterrotating $f$-modes from Ref.~\cite{Kruger:2019zuz}. The data points on both plots represent the corresponding $f$-mode frequencies measured from our simulations. Note that the section of the spectra containing the corotating modes at higher frequencies is not shown.}
    \label{fig:decspin_fmode_spectrum}
\end{figure}

We observe that the reduction of the NS spin magnitude caused by the resonance leads to a shift of the counterrotating mode to higher frequencies. This means that a feedback loop is possible: the overall increasing GW frequency approaches the tidal resonance, the NS gains angular momentum from the orbit and, therefore, spins down, the frequency of the counterrotating $f$-mode shifts upwards, and the GW frequency approaches the resonance again. Depending on the timescales of these processes, the resonance can be triggered multiple times or simply persist. This process is known as resonance locking~\cite{Witte:1999tq,Burkart:2012sq,Burkart:2013jua,Legred:2026tpz}.

In Fig.~\ref{fig:decspin_fmode_spectrum}, we illustrate this resonance locking by showing the GW spectra of isolated NSs with different spins $\chi \in \{0,0.25,0.375,0.5\}$ perturbed to excite $f$-mode oscillations. The resulting counterrotating $f$-mode frequencies closely follow the quasiuniversal relation of Ref.~\cite{Kruger:2019zuz}. For the NS in a binary, the GW frequency increases, while the resonance decreases the NS spin, and thus the counterrotating $f$-mode frequency is continuously pushed higher.

\begin{figure}[tp]
    \centering
    \includegraphics[width=1\linewidth]{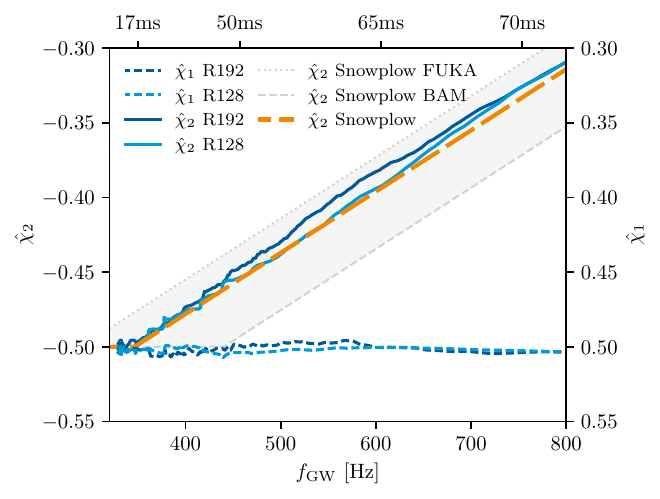}
    \caption{Spin measures $\hat{\chi}$ of each NS as a function of GW frequency $f_\mathrm{GW}$ at two resolutions, alongside the snowplow model for the resonating NS. The resonance locking begins at $f_\mathrm{GW} \lesssim 346$~Hz, with the resulting change of the resonating NS spin. Additionally, we plot the snowplow model prediction using the $f$-mode frequency calculated using the initial spin angular velocities from the initial data (\textsc{FUKA}) and from the evolution grid (\textsc{BAM}). The ticks on the top axis indicate the corresponding simulation times.}
    \label{fig:decspin_spins_freq}
\end{figure}

We consider a simple model where once the resonance is approached and the resonance locking sets in, the counterrotating $f$~mode increases to always maintain the resonance condition. We refer to this regime as the snowplow regime, and in turn, the model as the snowplow model. We assume that the tidal locking occurs at the frequency $f_\mathrm{lock}$, lying at a small constant offset $\Delta \sigma$ from the resonance frequency, i.e., $f_\mathrm{lock} \coloneq \sigma_i^\mathrm{u} - \Delta \sigma$.
 The quasiuniversal relation of Ref.~\cite{Kruger:2019zuz} relates the counterrotating mode frequency, $\sigma_i^\mathrm{u}$, to the spin angular frequency of the NS,~$\Omega$,

\begin{equation}
    \frac{\sigma_i^\mathrm{u}}{\sigma_0} = 1 + a_1^\mathrm{u} \left( \frac{|\Omega|}{\sigma_0}\right) + a_2^\mathrm{u} \left( \frac{|\Omega|}{\sigma_0}\right)^2,
\end{equation}
where  $\sigma_0$ is the $f$-mode frequency for the nonspinning NS, and  $a_1^\mathrm{u}=-0.193$ and $a_2^\mathrm{u} = -0.0294$.
Here, we can neglect the quadratic term, as $|\Omega/\sigma_i^u|~\ll~|a_1^\mathrm{u}/a_2^\mathrm{u}| \sim 6.6$, and $|\Omega/\sigma_0| \lesssim~3$.

We then reexpress the evolution of $\Omega$ as a function of the GW frequency using the resonance locking condition,

\begin{equation}
    \Omega(f_\mathrm{GW} | f_\mathrm{GW} > f_\mathrm{lock,0} ) = \Omega_0 -\frac{1}{a_1^\mathrm{u}} (f_\mathrm{GW}-f_\mathrm{lock,0}),
\end{equation}
where $\Omega_0$ is the initial angular velocity of the NS and $f_\mathrm{lock,0} = \sigma_{\mathrm{i},0}^\mathrm{u} - \Delta \sigma$ is the initial locking frequency, i.e., at the onset of continuous spin change. Note that we changed the sign of the $a_1^\mathrm{u}$ term to account for the fact that $\Omega < 0$ for our counterrotating star. Therefore, for the dimensionless spin of the resonating NS, we have
\begin{equation}
    \label{eq:snowplow}
    \chi(f_\mathrm{GW} | f_\mathrm{GW} \geq f_\mathrm{lock,0} ) = \chi_{0} - \frac{\bar{I} m}{a_1^\mathrm{u}} {(f_\mathrm{GW}-f_\mathrm{lock,0})},
\end{equation}
where $\chi_0$ is the initial spin of the NS, $m$ is its mass, $\bar{I}$ is the dimensionless moment of inertia of the NS (which is related to the moment of inertia $I$ as $\bar{I}=I/m^3$), while $\bar{I} m$ has units of inverse frequency.

Now, we adopt the dimensionless moment of inertia $\bar{I} = 12.18$ calculated via a quasiuniversal relation of Ref.~\cite{Breu:2016ufb} from the compactness of a nonrotating NS of the same baryonic mass, selecting their most appropriate fit with $\chi=0.4$, as for our case $|\chi| \leq 0.5$.

In Fig.~\ref{fig:decspin_spins_freq}, we show the evolution of the dimensionless spin $\hat{\chi}_2$ with GW frequency. At the onset of resonance, the spin changes linearly with the GW frequency, with the same slope for the two resolutions. In the same Fig.~\ref{fig:decspin_spins_freq}, we plot the prediction of the simple snowplow model. The evolution of the spin is predicted remarkably well. The low-resolution run shows the best agreement, while for the high-resolution run, the model deviates slightly. We suggest that lower numerical dissipation in the high-resolution simulation allows better resolution of angular momentum transfer, and, as a consequence, makes the frequency of the counterrotating mode shift faster than the GW frequency. In that case, the NS would periodically leave the resonance only to reenter it shortly after. This behavior appears to be supported by the staircase-like behavior at the beginning of the spin change.

Additionally, we plot the snowplow model predictions with  $f_\mathrm{lock,0}=\sigma_{\mathrm{i},0}^\mathrm{u}$, with $\sigma_{\mathrm{i},0}^\mathrm{u}$ calculated via the quasiuniversal relation of Ref.~\cite{Kruger:2019zuz} using initial spin angular frequency of the NS in the initial data reported by \textsc{FUKA}, and measured in \textsc{BAM} for the NS evolved in isolation after initial relaxation. The initial data has residual differential rotation (likely due to neglecting the second-order terms in spin), which relaxes to uniform rotation with a reduced $\Omega$ when evolved in \textsc{BAM}. The resonance frequency in \textsc{BAM} lies above the observed $f_\mathrm{lock,0}$, consistent with the expectation that resonance effects set in before reaching the resonance frequency.

\paragraph{\textbf{Outlook---}}
In this \textit{Letter}, we have shown, for the first time, that $f$-mode tidal resonance of one NS in an initially quasicircular BNS system is capable of introducing eccentricity in the late inspiral. In previous simulations of a fully symmetric NS binary~\cite{Kuan:2024jnw}, the tidal resonance happens for both NSs simultaneously and preserves quasicircularity. Therefore, in our simulation, the introduction of eccentricity can be attributed to the asymmetry of the system.

Introduction of eccentricity during late inspiral remains unmodelled by waveform models, and could bias conclusions of parameter estimation of real signals. 
For example, if the induced eccentricity is measured during late inspiral, and tidal resonance is not accounted in the analysis, one would conclude that the system had eccentricity $e_\mathrm{20Hz} \sim 0.1$ at 20~Hz, despite the system being quasicircular at that frequency. Therefore, detection of late-inspiral eccentricity would not necessarily point to a large natal eccentricity of a binary.

Additionally, we find that a simple model of tidal resonance locking is able to explain the evolution of the resonating NS spin as a function of the GW frequency remarkably well.
The tidal resonance locking effect arises from a continuous shift of the resonating mode frequency due to continuous spindown of the resonating NS. Further numerical simulations with different binary parameters would be required to assess the robustness of the model. For instance, systems with a low-mass NS or unequal mass ratios could be of particular interest.
The effect of tidal resonance locking could provide additional constraints on the behavior of matter at supranuclear densities, as it would reveal information about the $f$-mode oscillations in NSs. However, BNS systems with sufficiently large retrograde spins are unlikely to form under typical conditions. 

Current BNS waveform models show large disagreement with numerical-relativity data for highly spinning systems~\cite{Kuan:2025bzu}. This indicates that NS spin effects are either insufficiently modeled or entirely absent (cf. Refs.~\cite{Yu:2024uxt,Yu:2025ptm} for recent developments). The resonance-induced eccentricity, spin reduction, and tidal resonance locking reported here are examples of such effects.
Accounting for them in waveform models would allow a better description of such resonating BNS systems and enable unbiased parameter estimation. The resemblance of environmental effects (generally entering the GW phase at negative post-Newtonian orders~\cite{Barausse:2014tra,CanevaSantoro:2023aol}) by the tidal resonance locking suggests similar approaches might be useful to model it. 

\paragraph{\textbf{Acknowledgments---}}
We thank Anna Neuweiler for implementing the spin measure in \textsc{BAM} and for fruitful discussions. We thank Bernd Brügmann and Henrique Leonhard Gieg for useful discussions.
I.~M. and T.~D. gratefully acknowledge support from the Deutsche Forschungsgemeinschaft (DFG) under the project 504148597 (DI 2553/7). T.~D. acknowledges funding from the European Union (ERC, SMArt, 101076369).
The authors gratefully acknowledge the computing time granted by the Resource Allocation Board and provided on the supercomputer Emmy at NHR-Nord@Göttingen as part of the NHR infrastructure. Most calculations for this research were conducted with computing resources under the project bbp00049. 
Part of the simulations were performed on the DFG-funded research cluster Jarvis at the University of Potsdam (INST 336/173-1; project number: 502227537).
Views and opinions expressed are those of the authors only and do not necessarily reflect those of the European Union or the European Research Council. Neither the European Union nor the granting authority can be held responsible for them. 

\paragraph{\textbf{AI statement---}}
No large language models (LLMs) were used to produce code, text, figures, results, or models presented. LLMs were used to assist with identifying typographical errors.

\appendix

\section{High-frequency contamination}
As in Ref.~\cite{Kuan:2024jnw,Kuan:2025bzu}, the initial data for highly spinning NSs has inaccuracies that induce a high-frequency perturbation, polluting the waveform and other timeseries starting around $2660$~Hz and peaking at $3282$~Hz.
For that reason, we apply a low-pass filter with the cutoff frequency of $2660$~Hz to the GW frequency and spin time series.

\section{Influence of the Riemann solver choice}
\begin{figure}
    \centering
    \includegraphics[width=1\linewidth]{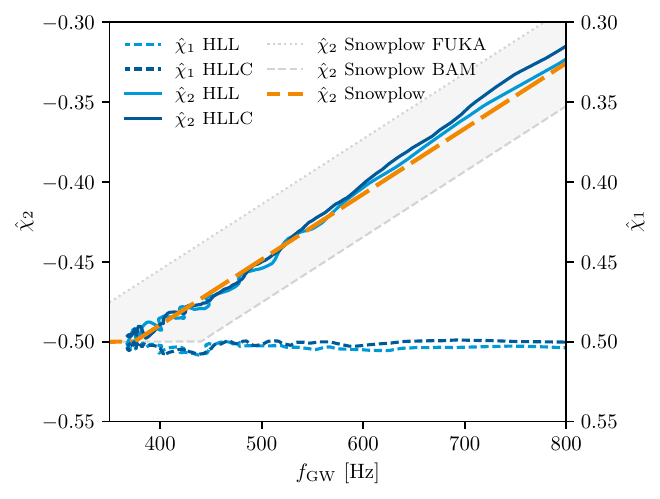}
    \caption{Same as Fig.~\ref{fig:decspin_spins_freq} but for different Riemann solvers employed, HLL and HLLC.}
    \label{fig:decspin_spins_freq_hllc}
\end{figure}

In Ref.~\cite{Kuan:2024jnw}, the authors highlighted the importance of employing Harten-Lax-van Leer-contact (HLLC) in contrast to Harten-Lax-van Leer (HLL) Riemann solver to resolve the tidal resonance and to obtain convergent results. Yet, their simulations differ not only in Riemann solvers but also in grid structure, namely, cell-centered and vertex-centered grids. It is possible that the vertex-centered grid (together with the HLL Riemann solver) can make the numerical scheme more diffusive.

Here, we perform additional simulations using the HLL and HLLC Riemann solvers. Unlike our main simulations, the binary configurations here have a higher initial orbital frequency of $180$~Hz, and are evolved at medium resolution R144 ($193$m grid spacing). In all our cases, the grid remains cell-centered. 

We show the results in Fig.~\ref{fig:decspin_spins_freq_hllc}. There, both HLL and HLLC produce nearly identical results. We do not observe the two-stage resonance in the evolution of the spin reported in Ref.~\cite{Kuan:2024jnw}.  Unlike in our main simulations, the resonance locking occurs at a higher frequency, $\sim 380~\mathrm{Hz}$. We attribute this to the fact that the initial data was produced closer to the resonance, therefore preventing the tidal resonance from developing earlier. 
Despite this, the snowplow model still describes the evolution of the spin remarkably well with the same slope as in the main text.
At the beginning of the resonance locking, the spin change has a periodic, staircase-like behavior. It can be explained by periodic reentry to the resonance if the frequency of the counterrotating $f$ mode increases more rapidly than the GW frequency. In that case, after leaving the resonance, the GW frequency would catch up with the $f$-mode frequency and reenter resonance locking.
We conclude that both Riemann solvers capture the same physical process of persistent tidal resonance, with minor accuracy differences.

\section{Data availability}
The data supporting the findings can be shared upon a reasonable request to the corresponding author.

An animation of the density evolution in the equatorial plane can be found at Ref.~\cite{markin_2026_24467100}.

\section{Carbon footprint}
\label{section:carbon_footprint}

The greenhouse gas (GHG) emissions, primarily from burning fossil fuels, are responsible for the ongoing, anthropogenically induced climate change~\cite{Oreskes_2004,Doran_2009,Archer:2009,Unger:2010,Cook_2013,Cook_2016,Fahey:2017,Lynas_2021,Myers_2021}. In the case of large simulations like ours, the GHG emissions can be drastically reduced by sourcing the electricity from renewables. For Emmy, the supercomputer at NHR-Nord@Göttingen, such a transition was done in 2021~\cite{NHRGoettingenPUE}.

Following our commitment to assess the climate impact of our work~\cite{Markin:2023fxx,Markin:2025oeo,Neuweiler:2025klw,Markin:2026eyc}, we report the estimated energy consumed and the corresponding GHG emissions produced by our simulations in Tab.~\ref{table:carbon_footprint}.

We collect and analyze the energy and emission data using the \textsc{Calcium} utility~\cite{calcium}. It acts as a process wrapper and collects the CPU model and usage metrics of its child processes. Then, \textsc{Calcium} automatically determines the CPU thermal design power, estimates the consumed energy, and uses the national electricity grid carbon intensity values to estimate the carbon emissions~\cite{owid_electricity_mix_2023}.

To account for the additional energy used by non-CPU components, we employ a node factor of $1.3$ for Emmy. We adopt a value of $1.03$ for the power usage effectiveness of Emmy~\cite{NHRGoettingenPUE}. Because the energy data is not available for Jarvis, we use the same node factor as for Emmy, and the power usage effectiveness of $1.1$ as a currently typical value for data centers in Germany.

\begin{table}[!htbp]
    \centering
    \begin{tabular}{l|c|c}
        & Energy & Induced grid emissions,\\
        & MWh & t\ch{CO2}e \\
        \hline
Evolution & 4.6 & 1.5 \\
Eccentricity reduction & 1.9 & 0.6 \\
Isolated NSs${}^*$ & 0.2 & 0.1 \\
\hline
Evolution (Suppl.\ Material) & 2.2 & 0.7 \\
Ecc. reduction (Suppl.\ Material) & 5.0 & 1.7 \\
\hline \hline
Total & 14.0 & 4.6 \\
    \end{tabular}
    \caption{The energy consumed and the corresponding \ch{CO2}e grid emissions induced. Only the run marked with an asterisk was performed on Jarvis.}
    \label{table:carbon_footprint}
\end{table}

\bibliography{main}

@article{Yu:2025ptm,
    author = "Yu, Hang and Lau, Shu Yan",
    title = "{Effective-one-body model for coalescing binary neutron stars: Incorporating tidal spin and enhanced radiation from dynamical tides}",
    eprint = "2501.13064",
    archivePrefix = "arXiv",
    primaryClass = "gr-qc",
    doi = "10.1103/PhysRevD.111.084029",
    journal = "Phys. Rev. D",
    volume = "111",
    number = "8",
    pages = "084029",
    year = "2025"
}

@article{Read:2008iy,
    author = "Read, Jocelyn S. and Lackey, Benjamin D. and Owen, Benjamin J. and Friedman, John L.",
    title = "{Constraints on a phenomenologically parameterized neutron-star equation of state}",
    eprint = "0812.2163",
    archivePrefix = "arXiv",
    primaryClass = "astro-ph",
    doi = "10.1103/PhysRevD.79.124032",
    journal = "Phys. Rev. D",
    volume = "79",
    pages = "124032",
    year = "2009"
}

@article{LIGOScientific:2021qlt,
    author = "Abbott, R. and others",
    collaboration = "LIGO Scientific, KAGRA, VIRGO",
    title = "{Observation of Gravitational Waves from Two Neutron Star\textendash{}Black Hole Coalescences}",
    eprint = "2106.15163",
    archivePrefix = "arXiv",
    primaryClass = "astro-ph.HE",
    reportNumber = "LIGO Document P2000357",
    doi = "10.3847/2041-8213/ac082e",
    journal = "Astrophys. J. Lett.",
    volume = "915",
    number = "1",
    pages = "L5",
    year = "2021"
}

@article{Papenfort:2021hod,
    author = "Papenfort, L. Jens and Tootle, Samuel D. and Grandcl\'ement, Philippe and Most, Elias R. and Rezzolla, Luciano",
    title = "{New public code for initial data of unequal-mass, spinning compact-object binaries}",
    eprint = "2103.09911",
    archivePrefix = "arXiv",
    primaryClass = "gr-qc",
    doi = "10.1103/PhysRevD.104.024057",
    journal = "Phys. Rev. D",
    volume = "104",
    number = "2",
    pages = "024057",
    year = "2021"
}

@article{Grandclement:2009ju,
    author = "Grandclement, Philippe",
    title = "{Kadath: A Spectral solver for theoretical physics}",
    eprint = "0909.1228",
    archivePrefix = "arXiv",
    primaryClass = "gr-qc",
    doi = "10.1016/j.jcp.2010.01.005",
    journal = "J. Comput. Phys.",
    volume = "229",
    pages = "3334--3357",
    year = "2010"
}

@article{Abac:2023ujg,
    author = "Abac, Adrian and Dietrich, Tim and Buonanno, Alessandra and Steinhoff, Jan and Ujevic, Maximiliano",
    title = "{New and robust gravitational-waveform model for high-mass-ratio binary neutron star systems with dynamical tidal effects}",
    eprint = "2311.07456",
    archivePrefix = "arXiv",
    primaryClass = "gr-qc",
    doi = "10.1103/PhysRevD.109.024062",
    journal = "Phys. Rev. D",
    volume = "109",
    number = "2",
    pages = "024062",
    year = "2024"
}

@article{Bruegmann:2006ulg,
    author = "Bruegmann, Bernd and Gonzalez, Jose A. and Hannam, Mark and Husa, Sascha and Sperhake, Ulrich and Tichy, Wolfgang",
    title = "{Calibration of Moving Puncture Simulations}",
    eprint = "gr-qc/0610128",
    archivePrefix = "arXiv",
    doi = "10.1103/PhysRevD.77.024027",
    journal = "Phys. Rev. D",
    volume = "77",
    pages = "024027",
    year = "2008"
}

@article{Thierfelder:2011yi,
    author = "Thierfelder, Marcus and Bernuzzi, Sebastiano and Bruegmann, Bernd",
    title = "{Numerical relativity simulations of binary neutron stars}",
    eprint = "1104.4751",
    archivePrefix = "arXiv",
    primaryClass = "gr-qc",
    doi = "10.1103/PhysRevD.84.044012",
    journal = "Phys. Rev. D",
    volume = "84",
    pages = "044012",
    year = "2011"
}

@article{Dietrich:2015iva,
    author = {Dietrich, Tim and Bernuzzi, Sebastiano and Ujevic, Maximiliano and Br\"ugmann, Bernd},
    title = "{Numerical relativity simulations of neutron star merger remnants using conservative mesh refinement}",
    eprint = "1504.01266",
    archivePrefix = "arXiv",
    primaryClass = "gr-qc",
    doi = "10.1103/PhysRevD.91.124041",
    journal = "Phys. Rev. D",
    volume = "91",
    number = "12",
    pages = "124041",
    year = "2015"
}

@article{Bernuzzi:2016pie,
    author = "Bernuzzi, Sebastiano and Dietrich, Tim",
    title = "{Gravitational waveforms from binary neutron star mergers with high-order weighted-essentially-nonoscillatory schemes in numerical relativity}",
    eprint = "1604.07999",
    archivePrefix = "arXiv",
    primaryClass = "gr-qc",
    doi = "10.1103/PhysRevD.94.064062",
    journal = "Phys. Rev. D",
    volume = "94",
    number = "6",
    pages = "064062",
    year = "2016"
}

@article{Dietrich:2018phi,
    author = {Dietrich, Tim and Radice, David and Bernuzzi, Sebastiano and Zappa, Francesco and Perego, Albino and Br\"ugmann, Bernd and Chaurasia, Swami Vivekanandji and Dudi, Reetika and Tichy, Wolfgang and Ujevic, Maximiliano},
    title = "{CoRe database of binary neutron star merger waveforms}",
    eprint = "1806.01625",
    archivePrefix = "arXiv",
    primaryClass = "gr-qc",
    doi = "10.1088/1361-6382/aaebc0",
    journal = "Class. Quant. Grav.",
    volume = "35",
    number = "24",
    pages = "24LT01",
    year = "2018"
}

@misc{calcium,
  author = {Ivan Markin},
  title = {Calcium: Tracking carbon footprint of computing},
  year = {2026},
  howpublished = {\url{https://codeberg.org/unkaktus/calcium}},
  note = {Version v1.10.0}
}

@article{Markin:2023fxx,
    author = "Markin, Ivan and Neuweiler, Anna and Abac, Adrian and Chaurasia, Swami Vivekanandji and Ujevic, Maximiliano and Bulla, Mattia and Dietrich, Tim",
    title = "{General-relativistic hydrodynamics simulation of a neutron star\textendash{}sub-solar-mass black hole merger}",
    eprint = "2304.11642",
    archivePrefix = "arXiv",
    primaryClass = "gr-qc",
    doi = "10.1103/PhysRevD.108.064025",
    journal = "Phys. Rev. D",
    volume = "108",
    number = "6",
    pages = "064025",
    year = "2023"
}

@article{Myers_2021,
doi = {10.1088/1748-9326/ac2774},
url = {https://dx.doi.org/10.1088/1748-9326/ac2774},
year = {2021},
month = {oct},
publisher = {IOP Publishing},
volume = {16},
number = {10},
pages = {104030},
author = {Myers, Krista F and Doran, Peter T and Cook, John and Kotcher, John E and Myers, Teresa A},
title = {Consensus revisited: quantifying scientific agreement on climate change and climate expertise among Earth scientists 10 years later},
journal = {Environmental Research Letters}
}

@article{Lynas_2021,
	title        = {Greater than 99\% consensus on human caused climate change in the peer-reviewed scientific literature},
	author       = {Lynas, Mark and Houlton, Benjamin Z and Perry, Simon},
	year         = 2021,
	month        = {oct},
	journal      = {Environmental Research Letters},
	publisher    = {IOP Publishing},
	volume       = 16,
	number       = 11,
	pages        = 114005,
	doi          = {10.1088/1748-9326/ac2966},
	url          = {https://dx.doi.org/10.1088/1748-9326/ac2966}
}

@article{Cook_2016,
doi = {10.1088/1748-9326/11/4/048002},
url = {https://dx.doi.org/10.1088/1748-9326/11/4/048002},
year = {2016},
month = {apr},
publisher = {IOP Publishing},
volume = {11},
number = {4},
pages = {048002},
author = {Cook, John and Oreskes, Naomi and Doran, Peter T and Anderegg, William R L and Verheggen, Bart and Maibach, Ed W and Carlton, J Stuart and Lewandowsky, Stephan and Skuce, Andrew G and Green, Sarah A and Nuccitelli, Dana and Jacobs, Peter and Richardson, Mark and Winkler, Bärbel and Painting, Rob and Rice, Ken},
title = {Consensus on consensus: a synthesis of consensus estimates on human-caused global warming},
journal = {Environmental Research Letters}
}

@article{Cook_2013,
doi = {10.1088/1748-9326/8/2/024024},
url = {https://dx.doi.org/10.1088/1748-9326/8/2/024024},
year = {2013},
month = {may},
publisher = {IOP Publishing},
volume = {8},
number = {2},
pages = {024024},
author = {Cook, John and Nuccitelli, Dana and Green, Sarah A and Richardson, Mark and Winkler, Bärbel and Painting, Rob and Way, Robert and Jacobs, Peter and Skuce, Andrew},
title = {Quantifying the consensus on anthropogenic global warming in the scientific literature},
journal = {Environmental Research Letters},
}

@article{Doran_2009,
author = {Doran, Peter T. and Zimmerman, Maggie Kendall},
title = {Examining the Scientific Consensus on Climate Change},
journal = {Eos, Transactions American Geophysical Union},
volume = {90},
number = {3},
pages = {22-23},
doi = {https://doi.org/10.1029/2009EO030002},
year = {2009}
}

@article{Oreskes_2004,
author = {Naomi Oreskes},
title = {The Scientific Consensus on Climate Change},
journal = {Science},
volume = {306},
number = {5702},
pages = {1686-1686},
year = {2004},
doi = {10.1126/science.1103618},
URL = {https://www.science.org/doi/abs/10.1126/science.1103618},
eprint = {https://www.science.org/doi/pdf/10.1126/science.1103618}}

@article{Kuan:2025bzu,
    author = "Kuan, Hao-Jui and Markin, Ivan and Ujevic, Maximiliano and Dietrich, Tim and Kiuchi, Kenta and Shibata, Masaru and Tichy, Wolfgang",
    title = "{The error budget of binary neutron star merger simulations for configurations with high spin}",
    eprint = "2506.02115",
    archivePrefix = "arXiv",
    primaryClass = "gr-qc",
    month = "6",
    journal = "",
    year = "2025"
}

@misc{NHRGoettingenPUE,
  author = {Gesellschaft für wissenschaftliche Datenverarbeitung mbH Göttingen},
  title = {{GWDG setzt auf klimafreundlichen IT-Betrieb}},
  howpublished = "\url{https://gwdg.de/about-us/press-releases/2021/press-release-4-2021/}",
  year = {2020}, 
}

@article{Douchin:2001sv,
    author = "Douchin, F. and Haensel, P.",
    title = "{A unified equation of state of dense matter and neutron star structure}",
    eprint = "astro-ph/0111092",
    archivePrefix = "arXiv",
    doi = "10.1051/0004-6361:20011402",
    journal = "Astron. Astrophys.",
    volume = "380",
    pages = "151",
    year = "2001"
}

@article{Planas:2025plq,
    author = "Planas, Maria de Lluc and Husa, Sascha and Ramos-Buades, Antoni and Valencia, Jorge",
    title = "{First eccentric inspiral-merger-ringdown analysis of neutron star-black hole mergers}",
    eprint = "2506.01760",
    archivePrefix = "arXiv",
    primaryClass = "astro-ph.HE",
    month = "6",
    year = "2025",
    journal = ""
}

@article{Jan:2025fps,
    author = "Jan, Aasim and Tsao, Bing-Jyun and O'Shaughnessy, Richard and Shoemaker, Deirdre and Laguna, Pablo",
    title = "{GW200105: A detailed study of eccentricity in the neutron star-black hole binary}",
    eprint = "2508.12460",
    archivePrefix = "arXiv",
    primaryClass = "gr-qc",
    month = "8",
    year = "2025",
    journal = ""
}

@article{Kacanja:2025kpr,
    author = "Kacanja, Keisi and Soni, Kanchan and Nitz, Alexander Harvey",
    title = "{Eccentricity signatures in LIGO-Virgo-KAGRA's BNS and NSBH binaries}",
    eprint = "2508.00179",
    archivePrefix = "arXiv",
    primaryClass = "gr-qc",
    month = "7",
    year = "2025",
    journal = ""
}

@article{Markin:2026eyc,
    author = "Markin, Ivan and Bulla, Mattia and Dietrich, Tim",
    title = "{Numerical simulations of black hole-neutron star mergers with equal and near-equal mass ratios}",
    eprint = "2601.19405",
    archivePrefix = "arXiv",
    primaryClass = "astro-ph.HE",
    month = "1",
    year = "2026",
    journal = "",
}

@article{Kuan:2024jnw,
    author = "Kuan, Hao-Jui and Kiuchi, Kenta and Shibata, Masaru",
    title = "{Tidal Resonance in Binary Neutron Star Inspirals: A High-Precision Study in Numerical Relativity}",
    eprint = "2411.16850",
    archivePrefix = "arXiv",
    primaryClass = "hep-ph",
    doi = "10.1103/j3zk-z17h",
    journal = "Phys. Rev. Lett.",
    volume = "135",
    number = "14",
    pages = "141403",
    year = "2025"
}

@article{Bethe:1998bn,
    author = "Bethe, Hans A. and Brown, G. E.",
    title = "{Evolution of binary compact objects which merge}",
    eprint = "astro-ph/9802084",
    archivePrefix = "arXiv",
    reportNumber = "SUNY-NTG-98-4",
    doi = "10.1086/306265",
    journal = "Astrophys. J.",
    volume = "506",
    pages = "780--789",
    year = "1998"
}

@article{Fragione:2018yrb,
    author = "Fragione, Giacomo and Grishin, Evgeni and Leigh, Nathan W. C. and Perets, Hagai. B. and Perna, Rosalba",
    title = "{Black hole and neutron star mergers in galactic nuclei}",
    eprint = "1811.10627",
    archivePrefix = "arXiv",
    primaryClass = "astro-ph.GA",
    doi = "10.1093/mnras/stz1651",
    journal = "Mon. Not. Roy. Astron. Soc.",
    volume = "488",
    number = "1",
    pages = "47--63",
    year = "2019"
}

@article{Yu:2024uxt,
    author = "Yu, Hang and Arras, Phil and Weinberg, Nevin N.",
    title = "{Dynamical tides during the inspiral of rapidly spinning neutron stars: Solutions beyond mode resonance}",
    eprint = "2404.00147",
    archivePrefix = "arXiv",
    primaryClass = "gr-qc",
    doi = "10.1103/PhysRevD.110.024039",
    journal = "Phys. Rev. D",
    volume = "110",
    number = "2",
    pages = "024039",
    year = "2024"
}

@article{Lai:1993di,
    author = "Lai, Dong",
    title = "{Resonant oscillations and tidal heating in coalescing binary neutron stars}",
    eprint = "astro-ph/9404062",
    archivePrefix = "arXiv",
    reportNumber = "CRSR-1064",
    doi = "10.1093/mnras/270.3.611",
    journal = "Mon. Not. Roy. Astron. Soc.",
    volume = "270",
    pages = "611",
    year = "1994"
}

@article{Kokkotas:1995xe,
    author = "Kokkotas, Kostas D. and Schaefer, Gerhard",
    title = "{Tidal and tidal resonant effects in coalescing binaries}",
    eprint = "gr-qc/9502034",
    archivePrefix = "arXiv",
    doi = "10.1093/mnras/275.2.301",
    journal = "Mon. Not. Roy. Astron. Soc.",
    volume = "275",
    pages = "301",
    year = "1995"
}

@article{Lai:1997wh,
    author = "Lai, Dong",
    title = "{Dynamical tides in rotating binary stars}",
    eprint = "astro-ph/9704132",
    archivePrefix = "arXiv",
    doi = "10.1086/304899",
    journal = "Astrophys. J.",
    volume = "490",
    pages = "847",
    year = "1997"
}

@article{Ho:1998hq,
    author = "Ho, Wynn C. G. and Lai, Dong",
    title = "{Resonant tidal excitations of rotating neutron stars in coalescing binaries}",
    eprint = "astro-ph/9812116",
    archivePrefix = "arXiv",
    doi = "10.1046/j.1365-8711.1999.02703.x",
    journal = "Mon. Not. Roy. Astron. Soc.",
    volume = "308",
    pages = "153",
    year = "1999"
}

@article{Lam:2022yeg,
    author = "Lam, Alan Tsz Lok and Shibata, Masaru and Kiuchi, Kenta",
    title = "{Numerical-relativity simulation for tidal disruption of white dwarfs by a supermassive black hole}",
    eprint = "2212.10891",
    archivePrefix = "arXiv",
    primaryClass = "astro-ph.HE",
    doi = "10.1103/PhysRevD.107.043033",
    journal = "Phys. Rev. D",
    volume = "107",
    number = "4",
    pages = "043033",
    year = "2023"
}

@article{Morras:2025xfu,
    author = "Morras, Gonzalo and Pratten, Geraint and Schmidt, Patricia",
    title = "{Orbital eccentricity in a neutron star - black hole binary merger}",
    eprint = "2503.15393",
    archivePrefix = "arXiv",
    primaryClass = "astro-ph.HE",
    reportNumber = "LIGO-DCC P2500105",
    doi = "10.3847/2041-8213/ae474c",
    journal = "Astrophys. J. Lett.",
    volume = "1000",
    number = "1",
    pages = "L2",
    year = "2026"
}

@article{Lara:2025kzj,
    author = "Lara, Guillermo and others",
    title = "{Signatures from metastable oppositely-charged black hole binaries in scalar Gauss-Bonnet gravity}",
    eprint = "2505.14785",
    archivePrefix = "arXiv",
    primaryClass = "gr-qc",
    month = "5",
    year = "2025",
    journal = ""
}

@article{Kruger:2019zuz,
    author = {Kr{\"u}ger, C. J. and Kokkotas, K. D.},
    title = "{Fast Rotating Relativistic Stars: Spectra and Stability without Approximation}",
    eprint = "1910.08370",
    archivePrefix = "arXiv",
    primaryClass = "gr-qc",
    doi = "10.1103/PhysRevLett.125.111106",
    journal = "Phys. Rev. Lett.",
    volume = "125",
    number = "11",
    pages = "111106",
    year = "2020"
}

@article{Samsing:2017xmd,
    author = "Samsing, Johan",
    title = "{Eccentric Black Hole Mergers Forming in Globular Clusters}",
    eprint = "1711.07452",
    archivePrefix = "arXiv",
    primaryClass = "astro-ph.HE",
    doi = "10.1103/PhysRevD.97.103014",
    journal = "Phys. Rev. D",
    volume = "97",
    number = "10",
    pages = "103014",
    year = "2018"
}

@article{Rodriguez:2017pec,
    author = "Rodriguez, Carl L. and Amaro-Seoane, Pau and Chatterjee, Sourav and Rasio, Frederic A.",
    title = "{Post-Newtonian Dynamics in Dense Star Clusters: Highly-Eccentric, Highly-Spinning, and Repeated Binary Black Hole Mergers}",
    eprint = "1712.04937",
    archivePrefix = "arXiv",
    primaryClass = "astro-ph.HE",
    doi = "10.1103/PhysRevLett.120.151101",
    journal = "Phys. Rev. Lett.",
    volume = "120",
    number = "15",
    pages = "151101",
    year = "2018"
}

@article{DiCarlo:2019pmf,
    author = "Di Carlo, Ugo N. and Giacobbo, Nicola and Mapelli, Michela and Pasquato, Mario and Spera, Mario and Wang, Long and Haardt, Francesco",
    title = "{Merging black holes in young star clusters}",
    eprint = "1901.00863",
    archivePrefix = "arXiv",
    primaryClass = "astro-ph.HE",
    doi = "10.1093/mnras/stz1453",
    journal = "Mon. Not. Roy. Astron. Soc.",
    volume = "487",
    number = "2",
    pages = "2947--2960",
    year = "2019"
}

@article{Banerjee:2016ths,
    author = "Banerjee, Sambaran",
    title = "{Stellar-mass black holes in young massive and open stellar clusters and their role in gravitational-wave generation}",
    eprint = "1611.09357",
    archivePrefix = "arXiv",
    primaryClass = "astro-ph.HE",
    doi = "10.1093/mnras/stw3392",
    journal = "Mon. Not. Roy. Astron. Soc.",
    volume = "467",
    number = "1",
    pages = "524--539",
    year = "2017"
}

@article{Zevin:2018kzq,
    author = "Zevin, Michael and Samsing, Johan and Rodriguez, Carl and Haster, Carl-Johan and Ramirez-Ruiz, Enrico",
    title = "{Eccentric Black Hole Mergers in Dense Star Clusters: The Role of Binary{\textendash}Binary Encounters}",
    eprint = "1810.00901",
    archivePrefix = "arXiv",
    primaryClass = "astro-ph.HE",
    reportNumber = "LIGO-P1800275",
    doi = "10.3847/1538-4357/aaf6ec",
    journal = "Astrophys. J.",
    volume = "871",
    number = "1",
    pages = "91",
    year = "2019"
}

@article{Banerjee:2017mgr,
    author = "Banerjee, Sambaran",
    title = "{Stellar-mass black holes in young massive and open stellar clusters and their role in gravitational-wave generation {\textendash} II}",
    eprint = "1707.00922",
    archivePrefix = "arXiv",
    primaryClass = "astro-ph.HE",
    doi = "10.1093/mnras/stx2347",
    journal = "Mon. Not. Roy. Astron. Soc.",
    volume = "473",
    number = "1",
    pages = "909--926",
    year = "2018"
}

@article{Heinze:2025usf,
    author = {Heinze, Felix M. and Br{\"u}gmann, Bernd and Dietrich, Tim and Markin, Ivan},
    title = "{Exploring the dynamics of general relativistic binary-single and binary-binary encounters of black holes}",
    eprint = "2505.03934",
    archivePrefix = "arXiv",
    primaryClass = "gr-qc",
    doi = "10.1103/2kgj-qkbn",
    journal = "Phys. Rev. D",
    volume = "112",
    number = "6",
    pages = "064068",
    year = "2025"
}

@article{Clarke:2026cuw,
    author = "Clarke, Teagan A. and Romero-Shaw, Isobel M. and Hoy, Charlie and Stegmann, Jakob and Lasky, Paul D. and Thrane, Eric",
    title = "{A universal framework to identify eccentric binary mergers: GW200105 case study}",
    eprint = "2605.18742",
    archivePrefix = "arXiv",
    primaryClass = "astro-ph.HE",
    month = "5",
    year = "2026",
    journal = ""
}

@ARTICLE{Reisenegger:1994,
       author = {{Reisenegger}, Andreas and {Goldreich}, Peter},
        title = "{Excitation of Neutron Star Normal Modes during Binary Inspiral}",
      journal = {\apj},
         year = 1994,
        month = may,
       volume = {426},
        pages = {688},
          doi = {10.1086/174105},
       adsurl = {https://ui.adsabs.harvard.edu/abs/1994ApJ...426..688R}
}

@article{Breu:2016ufb,
    author = "Breu, Cosima and Rezzolla, Luciano",
    title = "{Maximum mass, moment of inertia and compactness of relativistic stars}",
    eprint = "1601.06083",
    archivePrefix = "arXiv",
    primaryClass = "gr-qc",
    doi = "10.1093/mnras/stw575",
    journal = "Mon. Not. Roy. Astron. Soc.",
    volume = "459",
    number = "1",
    pages = "646--656",
    year = "2016"
}

@article{Peters:1964zz,
    author = "Peters, P. C.",
    title = "{Gravitational Radiation and the Motion of Two Point Masses}",
    doi = "10.1103/PhysRev.136.B1224",
    journal = "Phys. Rev.",
    volume = "136",
    pages = "B1224--B1232",
    year = "1964"
}

@article{Kowalska:2010qg,
    author = "Kowalska, I. and Bulik, T. and Belczynski, K. and Dominik, M. and Gondek-Rosinska, D.",
    title = "{The eccentricity distribution of compact binaries}",
    eprint = "1010.0511",
    archivePrefix = "arXiv",
    primaryClass = "astro-ph.CO",
    doi = "10.1051/0004-6361/201015777",
    journal = "Astron. Astrophys.",
    volume = "527",
    pages = "A70",
    year = "2011"
}

@article{Huerta:2013qb,
    author = "Huerta, E. A. and Brown, Duncan A.",
    title = "{Effect of eccentricity on binary neutron star searches in Advanced LIGO}",
    eprint = "1301.1895",
    archivePrefix = "arXiv",
    primaryClass = "gr-qc",
    reportNumber = "LIGO-DCC-1200187, KITP-NUMBER-NSF-KITP-12-194",
    doi = "10.1103/PhysRevD.87.127501",
    journal = "Phys. Rev. D",
    volume = "87",
    number = "12",
    pages = "127501",
    year = "2013"
}

@ARTICLE{1977A&A....57..383Z,
       author = {{Zahn}, J.-P.},
        title = "{Tidal friction in close binary systems.}",
      journal = "Astron. Astrophys.",
         year = 1977,
        month = may,
       volume = {57},
        pages = {383-394},
       adsurl = {https://ui.adsabs.harvard.edu/abs/1977A&A....57..383Z}
}

@article{Belczynski:2001uc,
    author = "Belczynski, Krzysztof and Kalogera, Vassiliki and Bulik, Tomasz",
    title = "{A Comprehensive study of binary compact objects as gravitational wave sources: Evolutionary channels, rates, and physical properties}",
    eprint = "astro-ph/0111452",
    archivePrefix = "arXiv",
    doi = "10.1086/340304",
    journal = "Astrophys. J.",
    volume = "572",
    pages = "407--431",
    year = "2001"
}

@ARTICLE{1961BAN....15..265B,
       author = {{Blaauw}, A.},
        title = "{On the origin of the O- and B-type stars with high velocities (the ``run-away'' stars), and some related problems}",
      journal = {Bulletin of the Astronomical Institutes of the Netherlands},
         year = 1961,
        month = may,
       volume = {15},
        pages = {265},
       adsurl = {https://ui.adsabs.harvard.edu/abs/1961BAN....15..265B}
}

@ARTICLE{1961BAN....15..291B,
       author = {{Boersma}, J.},
        title = "{Mathematical theory of the two-body problem with one of the masses decreasing with time}",
      journal = {Bulletin of the Astronomical Institutes of the Netherlands},
         year = 1961,
        month = may,
       volume = {15},
        pages = {291-301},
       adsurl = {https://ui.adsabs.harvard.edu/abs/1961BAN....15..291B}
}

@article{Armitage:2005xq,
    author = "Armitage, Philip J. and Natarajan, Priyamvada",
    title = "{Eccentricity of supermassive black hole binaries coalescing from gas rich mergers}",
    eprint = "astro-ph/0508493",
    archivePrefix = "arXiv",
    doi = "10.1086/497108",
    journal = "Astrophys. J.",
    volume = "634",
    pages = "921--928",
    year = "2005"
}

@article{Pfeiffer:2007yz,
    author = "Pfeiffer, Harald P. and Brown, Duncan A. and Kidder, Lawrence E. and Lindblom, Lee and Lovelace, Geoffrey and Scheel, Mark A.",
    editor = "Campanelli, Manuela and Rezzolla, Luciano",
    title = "{Reducing orbital eccentricity in binary black hole simulations}",
    eprint = "gr-qc/0702106",
    archivePrefix = "arXiv",
    doi = "10.1088/0264-9381/24/12/S06",
    journal = "Class. Quant. Grav.",
    volume = "24",
    pages = "S59--S82",
    year = "2007"
}

@ARTICLE{1983ApJS...53...73L,
       author = {{Lindblom}, L. and {Detweiler}, S.~L.},
        title = "{The quadrupole oscillations of neutron stars.}",
      journal = {The Astrophysical Journal Supplement Series},
         year = 1983,
        month = sep,
       volume = {53},
        pages = {73-92},
          doi = {10.1086/190884},
       adsurl = {https://ui.adsabs.harvard.edu/abs/1983ApJS...53...73L}
}

@article{Andersson:1997rn,
    author = "Andersson, Nils and Kokkotas, Kostas D.",
    title = "{Towards gravitational wave asteroseismology}",
    eprint = "gr-qc/9711088",
    archivePrefix = "arXiv",
    reportNumber = "AUTH-GRAV-97-03",
    doi = "10.1046/j.1365-8711.1998.01840.x",
    journal = "Mon. Not. Roy. Astron. Soc.",
    volume = "299",
    pages = "1059--1068",
    year = "1998"
}

@article{Ghosh:2025ban,
    author = "Ghosh, Rajes and Prasad, R. and Chakravarti, Kabir and Kumar, Prayush",
    title = "{Generalized perturbed Kepler problem: Gravitational wave imprints from eccentric compact binaries}",
    eprint = "2508.06245",
    archivePrefix = "arXiv",
    primaryClass = "gr-qc",
    doi = "10.1103/7mqv-rhc6",
    journal = "Phys. Rev. D",
    volume = "113",
    number = "8",
    pages = "084018",
    year = "2026"
}

@ARTICLE{Lidov:1962,
       author = {{Lidov}, M.~L.},
        title = "{The evolution of orbits of artificial satellites of planets under the action of gravitational perturbations of external bodies}",
      journal = {Planetary and Space Science},
         year = 1962,
        month = oct,
       volume = {9},
       number = {10},
        pages = {719-759},
          doi = {10.1016/0032-0633(62)90129-0},
       adsurl = {https://ui.adsabs.harvard.edu/abs/1962P&SS....9..719L}
}

@ARTICLE{Kozai:1962,
       author = {{Kozai}, Yoshihide},
        title = "{Secular perturbations of asteroids with high inclination and eccentricity}",
      journal = {The Astronomical Journal},
         year = 1962,
        month = nov,
       volume = {67},
        pages = {591-598},
          doi = {10.1086/108790},
       adsurl = {https://ui.adsabs.harvard.edu/abs/1962AJ.....67..591K}
}

@article{Zeipel:1909,
    author = {Zeipel, H. V.},
    title = {Sur l'application des séries de M. Lindstedt à l'étude du mouvement des comètes périodiques},
    journal = {Astronomische Nachrichten},
    volume = {183},
    number = {22-24},
    pages = {345-418},
    doi = {https://doi.org/10.1002/asna.19091832202},
    url = {https://onlinelibrary.wiley.com/doi/abs/10.1002/asna.19091832202},
    eprint = {https://onlinelibrary.wiley.com/doi/pdf/10.1002/asna.19091832202},
    year = {1909}
}

@ARTICLE{Robe:1968,
       author = {{Robe}, H.},
        title = "{{\'E}quilibre et oscillations des cylindres polytropiques compressibles en rotation}",
      journal = {Annales d'Astrophysique},
         year = 1968,
        month = feb,
       volume = {31},
        pages = {549},
       adsurl = {https://ui.adsabs.harvard.edu/abs/1968AnAp...31..549R}
}

@ARTICLE{Ledoux:1951,
       author = {{Ledoux}, P.},
        title = "{The Nonradial Oscillations of Gaseous Stars and the Problem of Beta Canis Majoris.}",
      journal = {The Astrophysical Journal},
         year = 1951,
        month = nov,
       volume = {114},
        pages = {373},
          doi = {10.1086/145477},
       adsurl = {https://ui.adsabs.harvard.edu/abs/1951ApJ...114..373L}
}

@article{Gaertig:2008uz,
    author = "Gaertig, Erich and Kokkotas, Kostas D.",
    title = "{Oscillations of rapidly rotating relativistic stars}",
    eprint = "0809.0629",
    archivePrefix = "arXiv",
    primaryClass = "gr-qc",
    doi = "10.1103/PhysRevD.78.064063",
    journal = "Phys. Rev. D",
    volume = "78",
    pages = "064063",
    year = "2008"
}

@article{Cuadra:2008xn,
    author = "Cuadra, J. and Armitage, P. J. and Alexander, R. D. and Begelman, M. C.",
    title = "{Massive black hole binary mergers within sub-pc scale gas discs}",
    eprint = "0809.0311",
    archivePrefix = "arXiv",
    primaryClass = "astro-ph",
    doi = "10.1111/j.1365-2966.2008.14147.x",
    journal = "Mon. Not. Roy. Astron. Soc.",
    volume = "393",
    pages = "1423",
    year = "2009"
}

@article{Markin:2025oeo,
    author = "Markin, Ivan and Puecher, Anna and Bulla, Mattia and Dietrich, Tim",
    title = "{Challenging a binary neutron star merger interpretation of GW230529}",
    eprint = "2508.08750",
    archivePrefix = "arXiv",
    primaryClass = "gr-qc",
    reportNumber = "LIGO-P2500457-v3",
    doi = "10.1103/6mtx-nftm",
    journal = "Phys. Rev. D",
    volume = "113",
    number = "2",
    pages = "024031",
    year = "2026"
}

@article{Neuweiler:2025klw,
    author = "Neuweiler, Anna and others",
    title = "{General-relativistic radiation magnetohydrodynamics simulations of binary neutron star mergers: The influence of spin on the multimessenger picture}",
    eprint = "2510.14850",
    archivePrefix = "arXiv",
    primaryClass = "astro-ph.HE",
    doi = "10.1103/mxlf-8sbm",
    journal = "Phys. Rev. D",
    volume = "113",
    number = "4",
    pages = "043038",
    year = "2026"
}

@inbook{Fahey:2017,
  author={Fahey, D. W. and Doherty, S. J. and Hibbard, K. A. and Romanou, A. and Taylor P. C.},
  editor={Wuebbles, D. J. and Fahey, D. W. and Hibbard, K. A. and Dokken, D. J. and Stewart, B. C. and Maycock, T. K.},
  title={Physical drivers of climate change},
  booktitle={Climate Science Special Report: Fourth National Climate Assessment, Volume I},
  year={2017},
  pages={73--113},
  publisher={U.S. Global Change Research Program},
  address={Washington, D.C.},
  doi={10.7930/J0513WCR},
}

@article{Archer:2009,
   author = "Archer, David and Eby, Michael and Brovkin, Victor and Ridgwell, Andy and Cao, Long and Mikolajewicz, Uwe and Caldeira, Ken and Matsumoto, Katsumi and Munhoven, Guy and Montenegro, Alvaro and Tokos, Kathy",
   title = "Atmospheric Lifetime of Fossil Fuel Carbon Dioxide", 
   journal= "Annual Review of Earth and Planetary Sciences",
   year = "2009",
   volume = "37",
   number = "Volume 37, 2009",
   pages = "117-134",
   doi = "https://doi.org/10.1146/annurev.earth.031208.100206",
   url = "https://www.annualreviews.org/content/journals/10.1146/annurev.earth.031208.100206",
   publisher = "Annual Reviews",
   issn = "1545-4495",
   type = "Journal Article",
  }

@article{Unger:2010,
author = {Nadine Unger  and Tami C. Bond  and James S. Wang  and Dorothy M. Koch  and Surabi Menon  and Drew T. Shindell  and Susanne Bauer },
title = {Attribution of climate forcing to economic sectors},
journal = {Proceedings of the National Academy of Sciences},
volume = {107},
number = {8},
pages = {3382-3387},
year = {2010},
doi = {10.1073/pnas.0906548107},
URL = {https://www.pnas.org/doi/abs/10.1073/pnas.0906548107},
eprint = {https://www.pnas.org/doi/pdf/10.1073/pnas.0906548107},
}

@misc{owid_electricity_mix_2023,
  author       = {Hannah Ritchie and Pablo Rosado and Max Roser},
  title        = {Carbon intensity of electricity},
  howpublished = {Our World in Data},
  year         = {2023},
  note         = {Data adapted from Ember, Energy Institute, Pinto et al., and Department for Energy Security and Net Zero of the UK. Retrieved from \url{https://ourworldindata.org/grapher/electricity-mix}},
  urldate      = {2026-09-30}
}

@article{Legred:2026tpz,
    author = "Legred, Isaac and Yunes, Nicol{\'a}s",
    title = "{Neutron Star--Primordial Black Hole Binaries: Dynamical Tides, Resonance Locking, and Gravitational Waves}",
    eprint = "2610.33314",
    archivePrefix = "arXiv",
    primaryClass = "gr-qc",
    month = "10",
    year = "2026",
    journal = "",
}

@article{Burkart:2013jua,
    author = "Burkart, Joshua and Quataert, Eliot and Arras, Phil",
    title = "{Dynamical resonance locking in tidally interacting binary systems}",
    eprint = "1312.4966",
    archivePrefix = "arXiv",
    primaryClass = "astro-ph.SR",
    doi = "10.1093/mnras/stu1366",
    journal = "Mon. Not. Roy. Astron. Soc.",
    volume = "443",
    number = "4",
    pages = "2957--2973",
    year = "2014"
}

@article{Witte:1999tq,
    author = "Witte, M. G. and Savonije, G. J.",
    title = "{Tidal evolution of eccentric orbits in massive binary systems: a study of resonance locking}",
    eprint = "astro-ph/9909073",
    archivePrefix = "arXiv",
    journal = "Astron. Astrophys.",
    volume = "350",
    pages = "129",
    year = "1999"
}

@article{Burkart:2012sq,
    author = "Burkart, Joshua and Quataert, Eliot and Arras, Phil and Weinberg, Nevin N.",
    title = "{Linear tides in inspiraling white dwarf binaries: resonance locks}",
    eprint = "1211.1393",
    archivePrefix = "arXiv",
    primaryClass = "astro-ph.SR",
    doi = "10.1093/mnras/stt726",
    journal = "Mon. Not. Roy. Astron. Soc.",
    volume = "433",
    pages = "332",
    year = "2013"
}

\end{document}